\documentclass[runningheads]{llncs}
\usepackage{graphicx}
\usepackage{cite}
\usepackage{listings}
\usepackage{pgfplots}
\pgfplotsset{compat=1.14}
\usepackage{subcaption}
\begin{document}
	\title{Recommending Scientific Videos based on Metadata Enrichment using Linked Open Data}
	\titlerunning{Recommending Scientific Videos using Linked Open Data}

	\author{Justyna Medrek$^{2}$ \and Christian Otto$^{1,3}$\orcidID{0000-0003-0226-3608} \and Ralph Ewerth$^{1,3}$\orcidID{0000-0003-0918-6297}}
	\authorrunning{Justyna Medrek \and Christian Otto \and Ralph Ewerth}
	\institute{Leibniz Information Centre for Science and Technology (TIB), Hannover, Germany, [firstname].[lastname]@tib.eu \and Leibniz Universit\"at Hannover, Germany, justa@mail.de \and L3S Research Center, Leibniz Universit\"at Hannover, Germany}
	\maketitle              % typeset the header of the contribution
	\begin{abstract}
		The amount of available videos in the Web has significantly increased not only for entertainment etc., but also to convey educational or scientific information in an effective way. There are several web portals that offer access to the latter kind of video material. One of them is the TIB AV-Portal of the Leibniz Information Centre for Science and Technology (TIB), which hosts scientific and educational video content. In contrast to other video portals, automatic audiovisual analysis (visual concept classification, optical character recognition, speech recognition) is utilized to enhance metadata information and semantic search. In this paper, we propose to further exploit and enrich this automatically generated information by linking it to the Integrated Authority File (GND) of the German National Library. This information is used to derive a measure to compare the similarity of two videos which serves as a basis for recommending semantically similar videos. A user study demonstrates the feasibility of the proposed approach.
		
		\keywords{video recommendation  \and semantic enrichment \and linked data}
	\end{abstract}
	
	\section{Introduction}
\label{sec:introduction}
 
Videos hold a great potential to communicate educational and scientific information. This is, for instance, reflected by e-Learning platforms such as Udacity (https://udacity.com) or Coursera (http://www.coursera.org). Another type of Web portals offers also access to scientific videos, one of them is the TIB AV-Portal (https://av.tib.eu) of the Leibniz Information Centre for Science and Technology (TIB).  Researchers can provide, search, and access scientific and educational audiovisual material, while benefiting from a number of advantages compared to other portals. First, submitted videos are reviewed to check whether they contain scientific or educational content. Second, videos are represented in a persistent way using DOIs (digital object identifier), potentially even at the segment and frame level, making it easy and reliable to reference them. Finally, audiovisual content analysis is applied in order to allow the user to not only search for terms in descriptive metadata (e.g., title, manually annotated keywords), but also in the audiovisual content, i.e., in the speech transcript, in the recognized overlaid or scene text through video OCR (optical character recognition), and keywords derived from visual concept and scene classification. 

Usually, recommender systems in online shopping platforms or video portals mainly rely on user-based information such as the viewing history~\cite{davidson2010youtube} or current trends~\cite{covington2016deep}. In this paper, we investigate the question how similar videos can be recommended based on their metadata, in particular, by additionally making use of automatically extracted metadata from audiovisual content analysis. This is relevant, for example, when users do not agree to track their search behavior or sufficient amount of user data is not available. Particularly, we propose to further exploit and enrich the entire set of available metadata, be it created manually or extracted automatically, in order to improve recommendations of semantically similar videos. In a first step, we utilize a Word2Vec approach~\cite{joulin2016bag} to make the semantic content of two videos comparable based on title, tags, and abstract. Then, the automatically extracted metadata about the audiovisual content is enriched by linking it to the Integrated Authority File (GND: Gemeinsame Normdatei) of the German National Library (DNB: Deutsche Nationalbibliothek). These two kinds of information are used to derive a measure to compare the content of two videos which serves as a basis for recommending similar video. A user study demonstrates the feasibility of the proposed approach. 

The paper is structured as follows. First, we give a brief overview of related work in Section 2. The proposed approach to generate video recommendations is presented in Section 3. Section 4 describes the conducted user study to evaluate the proposed approach, while Section 5 concludes the paper. 
\vspace{-0.1cm}
	\section{Related Work}
\label{sec:related_work}
\textit{Scientific Video Portals:} Yovisto is a scientific video portal that allows the user to search for information via text-based metadata~\cite{waitelonis2009augmenting, waitelonis2012towards}.  The users can reduce the number of search results by refining their query via additional criteria and grouping videos by language, organization, or category. On the contrary, to increase the scope of possible results, a tool for explorative search reveals interrelations between different types of videos in order to present a broader spectrum of results to the user. This is done by exploiting an ontology structure, which is part of every video element and Linked Open Data (LOD) resources, namely DBpedia (http://wiki.dbpedia.org).

Another similar portal is described by Marchionini~\cite{marchionini2006exploratory}, where the uploaded content is automatically fed into an automatic data analysis chain. Semantic entities are automatically assigned to each video segment resulting in a storyboard comprising the video content. In contrast to the AV-Portal, this information is hidden from the user. Marchionini's approach focuses on providing a good explorative search tool, i.e., a user should be able to find what s/he is looking for even when being unsure about the correct phrasing.
\textit{Recommendation Systems for Scientific Videos:} Clustering semantically similar videos is a possible approach to provide video recommendations based on a given, currently watched video. A fundamental problem of this research is the semantic gap between low-level features and high-level semantics portrayed in visual content~\cite{shaft1999nearest}. To circumvent this problem, textual cues can be used in addition to visual content. These can be manually added tags by the author of the video or automatically extracted keywords by machine learning algorithms. Either way, they are often superficial, noisy, incomplete or ambiguous which makes the process of clustering a challenge. Vahdat et al.~\cite{vahdat2014discovering} enrich the set of tags by modeling them from visual features and correct the existing ones by checking their agreement with the visual content. They are able to show that this method outperforms existing ones that use either modality and even the naive combination. 
Wang et al.~\cite{wang2016video} discover that by incorporating hierarchical information -- instead of considering a "flat" tag taxonomy -- the semantics of a video can be described even better. Despite only using two levels of abstraction in their hierarchical multi-label random forest model, strong correlations between ambiguous visual features and sparse, incomplete tags could be found.
\vspace{-0.1cm}
	\section{Enriching Video Metadata through Linked Open Data}
\label{sec:our_approach}
\begin{figure}[htbp]
	\centering
	\includegraphics[width=0.9\textwidth]{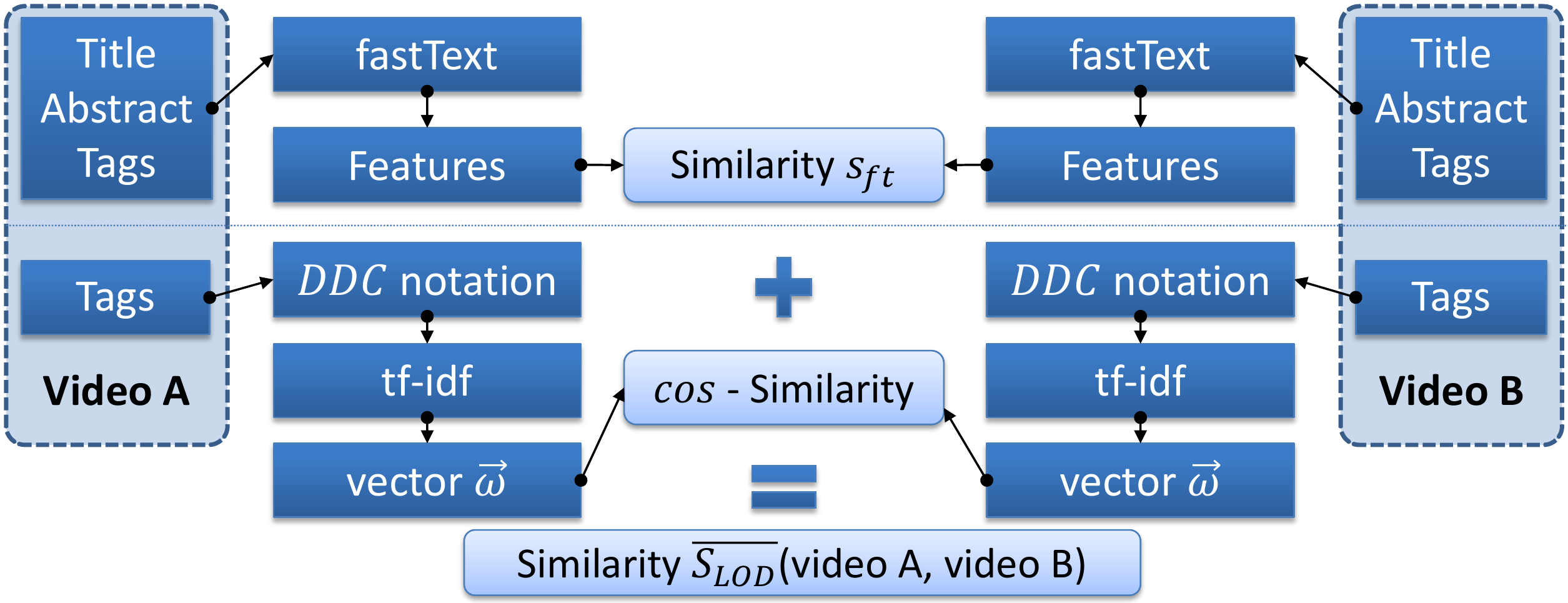}
	\caption{The general workflow of the approach combining the \textit{method without LOD} (upper half) with the features from the DDC notation (lower half).}
	\label{fig:workflow}
\end{figure}

In this section, we present our approach to enrich metadata with open data sources. First, the set of available metadata is described before the acquisition of additional information from an open data source is explained in Section~\ref{sec:linked_open_data}. Second, a similaritiy measure to compare videos based on a Word2Vec representation and enriched metadata is derived in Section~\ref{sec:similarity_measure}. The overall workflow is displayed in Figure~\ref{fig:workflow}.
The input of our system consists of manually generated and automatically extracted information, where the former comprises abstract and title. Additional inputs are the following automatically extracted \textbf{Tags} (see Figure~\ref{fig:workflow}) derived from: 1) Transcript based on speech recognition, 2) Results of video OCR, and 3) results of visual concept and scene classification. All of them have a representation in the German National Library, which is the key requirement for the enrichment process. 
\vspace{-0.3cm}
\subsection{Acquiring Additional Information from Open Data Source}
\label{sec:linked_open_data}
Automatically generated tags usually contain a certain amount of errors and noise. Although state-of-the-art algorithms can achieve human performance~\cite{shi2017end} in specific tasks and settings, issues with audio quality in lecture rooms or hardly legible handwritings can cause errors. We try to circumvent this problem by evaluating additional information provided by the German National Library. Besides information such as synonyms and related scientific publications, the \textit{Dewey Decimal Classification (DDC)} for every tag is provided. The DDC is a library classification system, which categorizes technical terms into ten classes via three-digit arabic numerals~\cite{reiner2008automatic}. These main classes are then further divided into subcategories denoted by the decimals after these three digits, where additional decimals depict a more specific subject. For instance, \textit{SPARQL} is contained in \textit{006.74 - Markup Language}, \textit{005.74 - Data files and Databases} and \textit{005.133 - Individual Programming Languages}, which yields valuable contextual information.
\vspace{-0.3cm}
\subsection{Defining a Similarity Measure for Scientific Videos}
\label{sec:similarity_measure}
Simply comparing two videos for mutual tags is not sufficient to determine semantic similarity. Even if two sets of tags have little to no overlap they might be highly correlated when their context is considered. We address this issue by utilizing fastText~\cite{joulin2016bag} to generate word embeddings, which has several advantages for this task. First, semantically similar words are modeled closer to one another so that a simple distance measure indicates the correlation of two words. Second, since fastText works on substrings rather than whole words it is able to produce valuable features even for misspelled or words unknown to the word embedding. Finally, a pre-trained model is available for a large number of languages. Title, tags, and abstract are taken from the metadata and processed via fastText. It generates a 300-dimensional feature vector for every word in the metadata. The average of these vectors is our representation for a particular video. This approach  is our baseline and denoted as \textit{method without LOD} in the sequel.

The improvement of this already powerful feature extraction method is the main contribution of this paper. It is achieved by incorporating the information provided by the DDC notation in addition to the fastText embeddings. As a preprocessing step we need to create a vector $\omega$, which consists of all DDC tags that occur in our dataset and which will be assigned to every video entry $v$. Since the upper level classes of the notation are also encoded in the codes of the classes at lower levels, we divide them accordingly. Therefore, the length of $\omega$ equals the total number of these tag fragments. For instance, if the video corpus would only contain the tags 005.74 and 005.133, we would split them into $5_1, 57_2, 51_2, 574_3, 513_3, 5133_4$ (indices mark the level in the hierarchy) resulting in a vector $\omega$ of length $6$. If a particular tag fragment occurs in a video, we set the corresponding bin in $\omega$ to the \textit{term frequency - inverse document frequency (tf-idf)}, or zero otherwise.  This assures that the more specific, and therefore more informative, DDC classes have more influence on the result. For example, if two tags share the main DDC class \textit{Science and Mathematics}, it does not mean that they are necessarily closely correlated, but if both share the class \textit{Data Compression} they most likely cover a similar topic. For the  "method with LOD" the two vectors $\omega_i$ and $\omega_j$ of video $v_i$ and $v_j$ are compared via cosine similarity. It is important to note that this method also uses the fastText features of the \textit{method without LOD}. In order to compute the overall similarity, both methods are applied and the average is used to form $s_{LOD}$ (see Figure \ref{fig:workflow}).
\vspace{-0.3cm}
	\section{Experimental Results}
\label{sec:experiments}

%\label{sec:prerequisites}
Videos of the TIB AV-Portal were used in the experiment. The complete stock of metadata that falls under the Creative Commons License CC0 1.0 Universal is made available by the TIB (https://av.tib.eu/opendata) as Resource Description Framework (RDF) triples. To extract the necessary annotations we utilized SPARQL. In a first step, it was necessary to keep only videos that allowed \textit{"derivate works"} in addition to the CC0 1.0 license, since content analysis is applied. $2\,066$ samples satisfied these conditions\footnote{as of June 16, 2017}. Unfortunately, word embeddings of two different languages cannot be directly compared forcing us to use a subset of videos with the same language (German in this case, 1\,430 videos). Annotations are represented in JSON format to make them easily accessible for future tasks without rebuilding the RDF graph. After gathering all tags of an entry, we employed another SPARQL query assigning a GND (German: \textit{Gemeinsame Normdatei}, English: Integrated Authority File) link to each tag, which is the key part of linking it to the data of the German National Library (DNB) and retrieving the corresponding DDC notations.

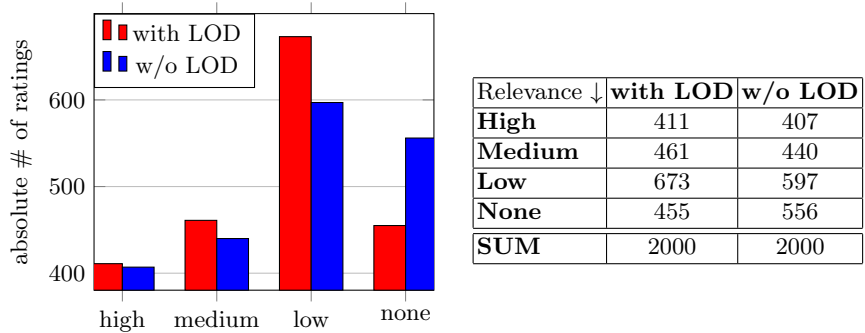
\begin{figure}[!ht]
	\centering
	\begin{subfigure}{0.5\textwidth}
		\begin{tikzpicture}
		\begin{axis}[
		ybar=0pt,
		width=\textwidth,
		bar width=12pt,
		symbolic x coords={high, medium, low, none},
		enlargelimits=0.1,
		legend style={at={(0,1)}, anchor=north west},
		ylabel={absolute \# of ratings},
		xtick=data,
		ymajorgrids=true]
		\addplot[fill=red] coordinates {
			(high, 411)
			(medium, 461)
			(low, 673)
			(none, 455)
		};
		\addplot[fill=blue] coordinates {
			(high, 407)
			(medium, 440)
			(low, 597)
			(none, 556)
		};
		\legend{with LOD, w/o LOD}
		\end{axis}
		\end{tikzpicture}
	\end{subfigure}%
	\begin{subfigure}{0.45\textwidth}
		\centering
		\begin{tabular}{| l | c | c |}%
			\hline
			Relevance $\downarrow$ & \textbf{with LOD} & \textbf{w/o LOD} \\ \hline
			\textbf{High}    & 411 & 407\\ \hline
			\textbf{Medium}  & 461 & 440\\ \hline
			\textbf{Low}     & 673 & 597\\ \hline
			\textbf{None}    & 455 & 556\\ \hline \hline
			\textbf{SUM}     & 2000 & 2000 \\ \hline
		\end{tabular}
	\end{subfigure}%
	\caption{Absolute number of votings for each relevance level in the user study.}
	\label{fig:results}
\end{figure}

We evaluated the quality of our similarity measure by conducting a user study with eight participants, five men and three women. A random selection of 50 videos was presented to every participant along with ten video recommendations, randomly either completely provided by the \textit{method without LOD} or the \textit{method with LOD}. The results were integrated by a Greasemonkey script in the Firefox browser. Every participant had to rate each of the ten recommendations from $0-3$, i.e., 0: not relevant; 1: low relevance; 2: medium relevance; 3: highly relevant. The results are displayed in Figure~\ref{fig:results}.
\newline
\\
The results show that the \textit{method with LOD} increases the number of video recommendations with medium ($4.56\%$) and low relevance ($11.29\%$), while the effect is small ($0.97\%$) for the highly relevant recommendations. However, the \textit{method with LOD} significantly decreases the number of irrelevant recommendations (by $18.17\%$). This indicates that this method is superior to the text-based method, most likely due to the hierarchical nature of the DDC notation. We assume that the rather small improvement for the very relevant recommendations is a result of the restrictions we had to oblige to (license and language), i.e., the relatively small set of remaining videos ($1\,430$) does not contain more highly relevant samples. A chi-square test shows that the \textit{method with LOD} is significantly better than our baseline (Chi-Square=$15.1471$, p-value=$0.001695$).
\vspace{-0.3cm}
	\section{Conclusions}
\label{sec:conclusion}

In this paper, we have proposed a method to generate recommendations for scientific videos based on noisy, automatically extracted tags by utilizing linked open data to weave in hierarchical semantic metadata. This enables users to find relevant information more quickly improving their overall learning experience. In future work, we plan to incorporate recommendations for scientific papers or definitions of technical terms through linked open data.
	
	\bibliographystyle{splncs04}
	\bibliography{1_main}
	
\end{document}